\documentstyle[jkas]{article}

\runningauthor {BYEONG-CHEOL LEE ET AL.}
\runningtitle{PLANETARY COMPANION IN K GIANT $\sigma$ PERSEI}

\month{April} \year{2014} \volume{xx} \issueno{x}
\beginpage{1}\endpage{9}
\begin{document}
\title{PLANETARY COMPANION IN K GIANT $\sigma$ PERSEI\thanks{Based on observations made with the Bohyunsan Observatory Echelle Spectrograph (BOES) instrument on the 1.8-m telescope at Bohyunsan Optical Astronomy Observatory in Korea.}}
\author{Byeong-Cheol Lee$^{1,2}$, Inwoo Han$^{1}$, Myeong-Gu Park$^{3}$, David E. Mkrtichian$^{4,5}$, Gwanghui Jeong$^{1,2}$, \\
Kang-Min Kim$^{1}$, and Gennady Valyavin$^{6}$}
\address{$^1$ Korea Astronomy and Space Science Institute 776, Daedeokdae-ro, Yuseong-gu, Daejeon 305-348, Korea \\
 {\it E-mail: bclee@kasi.re.kr, iwhan@kasi.re.kr, tlotv@kasi.re.kr, kmkim@kasi.re.kr}}
\address{$^2$ Astronomy and Space Science Major, University of Science and Technology, Gajeong-ro Yuseong-gu, Daejeon 305-333, Korea}
\address{$^3$ Department of Astronomy and Atmospheric Sciences, Kyungpook National University, Daegu 702-701, Korea\\
 {\it E-mail: mgp@knu.ac.kr}}
\address{$^4$ National Astronomical Research Institute of Thailand, Chiang Mai 50200, Thailand}
\address{$^5$ Crimean Astrophysical Observatory, Taras Shevchenko National University of Kyiv, 98409, Nauchny, Crimea, Ukraine\\
 {\it E-mail: davidmkrt@gmail.com}}
\address{$^6$ Special Astrophysical Observatory, Russian Academy of Sciences, Nizhnii Arkhyz, 369167, Russia\\
 {\it E-mail: gvalyavin@sao.ru}}
\address{\normalsize{\it (Received February 20, 2014; Accepted March 4, 2014)}}
%\offprints{Byeong-Cheol Lee}

%--------------------------------------------------------------------
\abstract{We report the detection of an exoplanet candidate in orbit around $\sigma$ Persei from a radial velocity (RV) survey. The system exhibits periodic RV variations of 579.8 $\pm$ 2.4 days. The purpose of the survey is to search for low-amplitude and long-period RV variations in giants and examine the origin of the variations using the fiber-fed Bohyunsan Observatory Echelle Spectrograph installed at the 1.8-m telescope of Bohyunsan Optical Astronomy Observatory in Korea. We present high-accuracy RV measurements of $\sigma$~Per made from December 2003 to January 2014. We argue that the RV variations are not related to the surface inhomogeneities but instead a Keplerian motion of the planetary companion is the most likely explanation. Assuming a stellar mass of 2.25 $\pm$ 0.5 $M_{\odot}$, we obtain a minimum planetary companion mass of 6.5 $\pm$ 1.0 $M_{\rm{Jup}}$, with an orbital semi-major axis of 1.8 $\pm$ 0.1 AU, and an eccentricity of 0.3 $\pm$ 0.1 around $\sigma$ Per.
}
\maketitle

%--------------------------------------------------------------------
\section{INTRODUCTION}
Ever since the discovery of the first exoplanet around a main-sequence (MS) star by Mayor $\&$
Queloz (1995) using the radial velocity (RV) method, approximately 1000 exoplanets have been found so
far. It has been demonstrated that the RV method is more efficient for late spectral type stars
having an ample number of spectral lines. Planet searches generally focus on late F, G, and K low-mass
dwarfs. Stellar and hence planet masses in such systems can be precisely determined.
The RV search for exoplanets around early class B and A-type stars is difficult due to the high-rotational velocities of these stars
and the low accuracy of RV measurements.

For giant stars, on the other hand, RV shifts can be measured with high precision due to their sharp spectral lines.
Lack of knowledge about planet formation makes RV surveys of giant stars an important endeavor.
Compared to MS stars, however, giants may show more complex RV variations due to the influence of various surface processes on the line profiles,
such as chromospheric activities, spots, large convection cells, and stellar pulsations.
It is thus difficult to identify the origin of RV variations in giant star systems. Also, it is difficult to determine
accurate masses  of the host giant stars, because giant stars of different masses are located in roughly
the same region of the H-R Diagram.
Finally, there appears to exist a selection bias in that only massive planets
with relatively long orbital periods have been detected around giant stars, whereas surveys of
MS stars tend to find low-mass planets with short orbital periods.

Frink et al. (2002) discovered the first planetary companion around the K giant star $\iota$ Dra (K2).
Since then, several companions around K giant stars have been detected through the RV
method. Recognizing the importance of exoplanet formation and evolution, several exoplanet
survey groups began exploring evolved K giants (Frink et al. 2002; Setiawan et al. 2003; Hatzes
et al. 2005; D{\"o}linger et al. 2007; Han~et~al. 2010).

We initiated a precise RV survey using the 1.8-m telescope at Bohyunsan Optical Astronomy Observatory
(BOAO) in 2003 to search for exoplanets and to study the  oscillations  for 6 F giants, 55 K
giants, and 10 M giants. In the course of the program we have reported the discovery of six new exoplanets (Han~et~al. 2010; Lee et al. 2011,
2012b, 2012c, 2013), an exoplanet candidate (Lee et al. 2012a), and confirmed the exoplanet around
Pollux (Han et al. 2008). The K giant $\sigma$ Per is one of the 55 K giants in our sample for which we have
obtained precise RV measurements in the past 11 years at BOAO. Here we report the detection
of low-amplitude and long-period RV variations for $\sigma$ Per, possibly caused by a
planetary companion. In Sect. 2, we describe the observations and data reduction. In Sect. 3, the
stellar characteristics of the host star are derived. The RV variation measurements and possible origins
are presented in Sect. 4. Finally, in Sect. 5, we discuss the main results from our study.

%--------------------------------------------------------------------
\section{OBSERVATIONS AND REDUCTION}
The observations were carried out as a part of the RV variation study of 55 K giants. We used the fiber-fed,
high-resolution ($\emph{R}$ = 90 000) Bohyunsan Observatory Echelle Spectrograph (BOES; Kim et al.
2007) installed at the 1.8-m telescope of BOAO, Korea. The BOES spectra covered the wavelength coverage
from 3500 to 10500~{\AA}. To provide precise RV measurements, an iodine absorption (I$_{2}$) cell
was used with a wavelength region of 4900$-$6000 {\AA}. The characteristic signal-to-noise (S/N) for the I$_{2}$ region was
about 200 at typical exposure times ranging from 5 to 15 minutes.

We recorded 71 spectra for $\sigma$ Per from December 2003 to January 2014 (52 nights in total).
The basic reduction of spectra was performed using the IRAF software package and the DECH (Galazutdinov
1992) code. The precise RV measurements related to the I$_2$ analysis were undertaken using the
RVI2CELL (Han~et~al. 2007) code, which is based on a method by Butler~et~al. (1996) and Valenti et al.
(1995). However, for the modeling of the instrument profile, we used the matrix formula described by
Endl et al. (2000). We solved the matrix equation using singular value decomposition instead of the
maximum entropy method adopted by Endl et al. (2000).

The long-term stability of the BOES was demonstrated by observing the RV standard star $\tau$ Ceti,
which was constant with an rms scatter of 6.8 m s$^{-1}$ (Lee et al. 2013). The RV measurements for
$\sigma$ Per are listed in Table \ref{tab1} and shown in Figure~\ref{orbit}.

%-------------------------------------------------------------
\begin{deluxetable}{rrrrrrrrr}
\tablecolumns{9}
\tablewidth{0pc}
\tablecaption{RV measurements for $\sigma$ Per from December 2003 to January 2014 using the BOES. \label{tab1}}
\tablehead{
\colhead{JD}  &
\colhead{$\Delta$RV} &
\colhead{$\pm \sigma$}&
\colhead{JD}&
\colhead{$\Delta$RV}&
\colhead{$\pm \sigma$}&
\colhead{JD}&
\colhead{$\Delta$RV}&
\colhead{$\pm \sigma$} \\
\colhead{ $-$2450000} &
\colhead{m\,s$^{-1}$} &
\colhead{m\,s$^{-1}$} &
\colhead{ $-$2450000} &
\colhead{m\,s$^{-1}$} &
\colhead{m\,s$^{-1}$} &
\colhead{ $-$2450000} &
\colhead{m\,s$^{-1}$} &
\colhead{m\,s$^{-1}$}
}
\startdata
2977.296077 &      126.9  &      6.8  &   3728.916722  &    $-$56.3  &      8.0  &   4752.336940  &       22.5   &     7.5  \\
2978.048193 &       92.8  &      6.2  &   3759.115965  &   $-$101.5  &      8.0  &   4755.355225  &     $-$3.2   &     7.9  \\
2978.055345 &       89.6  &      6.3  &   3761.032748  &    $-$85.1  &      7.2  &   4847.174601  &   $-$130.2   &     7.2  \\
2981.086172 &      113.2  &      6.0  &   3809.013764  &   $-$111.3  &      8.8  &   4879.983438  &    $-$96.9   &     6.7  \\
2981.092237 &      111.1  &      7.3  &   3818.967080  &    $-$19.7  &      6.8  &   4930.953826  &    $-$49.4   &     7.5  \\
3044.998373 &    $-$88.9  &      7.4  &   3818.980644  &    $-$19.6  &      7.0  &   5172.123983  &       30.1   &     7.1  \\
3045.006659 &    $-$91.0  &      7.0  &   3819.950976  &    $-$99.0  &      8.0  &   5251.069391  &       40.5   &     7.2  \\
3046.038607 &     $-$1.9  &      6.8  &   3820.988525  &    $-$80.0  &      8.3  &   5454.324212  &    $-$91.8   &     7.1  \\
3046.047009 &     $-$0.5  &      7.3  &   4038.348295  &      123.6  &      7.8  &   5554.054953  &    $-$34.9   &     8.0  \\
3048.038120 &    $-$39.9  &      7.4  &   4123.034383  &       40.6  &      7.2  &   5555.010664  &       10.3   &    19.3  \\
3048.044219 &    $-$32.5  &      7.4  &   4126.078753  &      102.3  &      7.9  &   5581.091335  &    $-$48.2   &     7.6  \\
3253.308818 &    $-$84.5  &      8.0  &   4166.951262  &       28.3  &      7.3  &   5842.326935  &       98.1   &     6.3  \\
3302.087308 &     $-$9.5  &      6.2  &   4209.932513  &     $-$1.3  &     13.4  &   5932.979094  &       18.3   &     6.8  \\
3302.093338 &    $-$12.8  &      6.5  &   4209.940314  &       13.0  &     12.9  &   5962.938553  &     $-$2.8   &     8.1  \\
3303.332572 &    $-$11.5  &      6.4  &   4396.316119  &    $-$30.4  &      7.2  &   6023.938532  &    $-$97.5   &     7.7  \\
3303.338209 &    $-$11.7  &      6.6  &   4469.976887  &       55.5  &      7.9  &   6176.321855  &     $-$9.7   &     7.5  \\
3303.343035 &    $-$12.6  &      6.8  &   4470.912387  &    $-$35.9  &      7.6  &   6257.989203  &    $-$12.3   &     8.7  \\
3303.347596 &    $-$13.4  &      7.3  &   4470.917514  &    $-$32.0  &      7.0  &   6376.950984  &       86.9   &     9.5  \\
3333.282921 &    $-$21.5  &      6.7  &   4470.922734  &    $-$34.8  &      7.4  &   6377.987203  &      132.3   &     8.6  \\
3356.173663 &       66.9  &      7.7  &   4470.927375  &    $-$35.8  &      7.6  &   6619.937210  &    $-$38.8   &     7.0  \\
3431.024416 &       68.8  &     10.0  &   4506.081969  &       42.8  &      7.0  &   6679.081040  &    $-$54.2   &     7.6  \\
3432.960068 &       65.3  &      9.2  &   4536.994775  &       95.7  &      8.8  &   6679.085716  &    $-$52.1   &     8.2  \\
3432.966734 &       66.9  &      7.5  &   4719.182078  &     $-$2.2  &      8.2  &   6679.090380  &    $-$52.4   &     8.1  \\
3433.099917 &       89.9  &     12.0  &   4726.342887  &     120.3   &      8.3  &                &              &          \\
\enddata
\end{deluxetable}
%-------------------------------------------------------------

\section{STELLAR CHARACTERISTICS}
Evolved stars are known to exhibit low-amplitude and long-period RV variations produced by
rotation and surface activity such as spots, plages, or filaments. The investigation of the stellar
characteristics thus is important in discerning the nature of any RV variations.

\subsection{FUNDAMENTAL PARAMETERS}
We assumed the main photometric parameters for $\sigma$~Per (= HD 21552 = HR 1052 = HIP 16335) from
the \emph{HIPPARCOS} catalog (ESA 1997). $\sigma$ Per is a cool evolved star of type K3 III and has an
apparent magnitude V~=~4.36. The astrometric parallax ($\pi$) was taken from the improved
\emph{HIPPARCOS} value by van Leeuwen (2007), 9.07 $\pm$ 0.26 mas.
Other stellar parameters are taken from Luck \& Heiter (2007): $T_{\mathrm{eff}}$ =
4367~K, [Fe/H]~= -- 0.22 $\pm$ 0.23, and log $\it g$ = 1.8~$\pm$~0.1. We also determined atmospheric
parameters directly from our spectra by measuring 274 equivalent widths (EW) of Fe~I and Fe~II
lines. We estimated $T_{\mathrm{eff}}$ = 4201~$\pm$~22~K, [Fe/H] = -- 0.21 $\pm$ 0.06, log $\it g$ =
1.77~$\pm$~0.09, and $v_{micro}$~= 1.5 $\pm$ 0.1 using the program TGVIT (Takeda et al. 2005).
Our estimated values were in agreement with Luck \& Heiter (2007). Table~\ref{tab2} summarizes the
basic stellar parameters of $\sigma$~Per.

\subsection{STELLAR MASS}
Due to disagreement in the mass of the host star as measured by different authors and methods,
the planet mass is subject to some uncertainty..
In order to derive stellar mass, one needs stellar parameters, such as the effective temperature.
Luck \& Heiter (2007) applied two basic methods of mass determination for nearby giants
including $\sigma$ Per: photometric and spectroscopic.
The photometric mass was derived using the isochrones of
Bertelli et al. (1994), interpolated using the absolute V magnitude and photometric temperature.
The authors then derived the luminosity and surface gravity using standard relations. The spectroscopic mass was
determined using the direct spectroscopic value for the surface gravity and the luminosity using
standard relations. Both mass estimates are shown in Table~\ref{tab2}. The mass derived from the
spectroscopic data is larger than that derived from the photometry. The difference is due to the
larger surface gravity measured by the spectroscopic analysis, which demands larger mass.
The use of different temperatures (i.e., photometric vs. spectroscopic) has little effect
on the physical determination of the mass. We considered results from both the photometric
and spectroscopic approaches.

%-------------------------------------------------------------
\begin{table}
\begin{center}
\caption{Stellar parameters for $\sigma$ Per assumed in the present paper
\label{tab2}}
\doublerulesep2.0pt
\renewcommand\arraystretch{1.0}
\begin{tabular}{lccc}
\hline
\hline
Parameter             & Unit & Value  & Ref. \\
\hline
    Spectral type      &     & K3 III   & 1  \\
    $\textit{$m_{v}$}$ &[mag]& 4.36     & 1  \\
    $\textit{$M_{v}$}$ &[mag]& $-$ 1.64     & 2\tablenotemark{b}  \\
    $\textit{B-V}$     &[mag]& 1.367 $\pm$ 0.004  & 1 \\
%    age                &[Gyr]& 2.0 $\pm$ 0.5  & 2  \\
    $\emph{d}$         &[pc] & 108.3          & 2  \\
    RV                 &[km s$^{-1}$]& 14.36 $\pm$ 0.19  & 2 \\
    $\pi$              &[mas]& 9.07 $\pm$ 0.26 &  3  \\
    $T_{\rm{eff}}$     &[K] & 4367 $\pm$ 100   &  2\tablenotemark{a}  \\
                       &    & 4201 $\pm$ 22    &  4  \\
    $\rm{[Fe/H]}$      &[dex]& $-$ 0.22 $\pm$ 0.23  & 2\tablenotemark{a}  \\
                       &     & $-$ 0.21 $\pm$ 0.06  & 4  \\
    log $\it g$        &[cgs]& 1.8  $\pm$ 0.1     &  2\tablenotemark{a}  \\
                       &     & 1.56               &  2\tablenotemark{b}  \\
                       &     & 1.77 $\pm$ 0.09 &  4  \\
    $v_{\rm{micro}}$        &[km s$^{-1}$]& 2.01 $\pm$ 0.3  & 2\tablenotemark{a}  \\
                            &             & 1.5  $\pm$ 0.1  &  4  \\
%    $v_{\rm{macro}}$        &[km s$^{-1}$]& 4.27     & 8   \\
    $\textit{$R_{\star}$}$  &[$R_{\odot}$]& 28 &  5  \\
    $\textit{$M_{\star}$}$  &[$M_{\odot}$]& 2.25 $\pm$ 0.5 &  2\tablenotemark{a}  \\
                            &             & 1.82           &  2\tablenotemark{b}  \\
    $\textit{$L_{\star}$}$  &[$L_{\odot}$] & 316.2    & 2\tablenotemark{a} \\
                            &              & 363.1    & 2\tablenotemark{b} \\
                            &              & 308.22   & 6  \\
    $v_{\rm{rot}}$ sin($i$)  &[km s$^{-1}$] & 1.0    &  7  \\
                            &              & 2.21   &  8  \\
                            &              & 1.3    &  4  \\
%   log $R^{'}_{\rm HK}$     &              &  --    & 12   \\
    $P_{\rm{rot}}$ / sin($i$)  &[days] & 641$-$1288 &  4 \\
\hline
\end{tabular}
\end{center}
\begin{tabnote}
\textbf{References.} 1 \emph{HIPPARCOS} (ESA 1997); 2\tablenotemark{a}~ Luck \& Heiter (2007) estimated by spectroscopic method; 2\tablenotemark{b}~ Luck \& Heiter (2007) estimated by photometric method; 3 van Leeuwen (2007); 4 This work; 5 Pasinetti-Fracassini et al. (2001); 6 McDonald et al. (2012); 7 de Medeiros \& Mayor (1999); 8 Hekker \& Mel{\'e}ndez (2007) \\
%2 Derived using the online tool (http://stevoapd.inaf.it/cgi-bin/param);
\end{tabnote}
\end{table}

\subsection{ROTATIONAL VELOCITY}
In evolved stars, low-amplitude and long-period RV variations may arise from the rotation (Lee et al. 2008, 2012a).
Thus, the rotational period is very important in distinguishing between RV variations and rotational modulation.
In order to estimate the stellar rotational velocity, we used a line-broadening model by Takeda et al. (2008).
The observed stellar spectrum was fitted by convolving the intrinsic spectrum model and a total macrobroadening function.
The intrinsic spectrum model can be computed if a model atmosphere characterized by the
microturbulence, $v_{\mathrm{micro}}$, and an elemental abundance are given. The total
macrobroadening function is a convolution of three component functions:
the instrumental broadening dispersion ($v_{\mathrm{ip}}$), the rotational broadening velocity
($v_{\mathrm{rot}}$), and the macroturbulence ($v_{\mathrm{macro}}$). These component functions broaden spectral lines
without altering their EW. Distinguishing between the broadening components is difficult for spectra
of slowly rotating late type stars because they have similar intrinsic line profiles.

For the determination of line broadening, we used the automatic spectrum-fitting technique (Takeda 1995) for the spectrum within the wavelength range  6080$-$ 6089 {\AA}. This technique employs seven free parameters that specify the best fitting solution: the abundances of six elements (Si, Ti, V, Fe, Co, and Ni) and the total macrobroadening parameter ($v_{\mathrm{M}}$) which is given as

\begin{equation}\label{eq:Dss}
    v^{2}_{\mathrm{M}} = v^{2}_{\mathrm{ip}} + v^{2}_{\mathrm{rot}} + v^{2}_{\mathrm{macro}},
\end{equation}

\hskip -15pt
where $v_{\mathrm{ip}}$ can be described as $(c/R)/({2\sqrt{\mathrm{ln}2}})$ with ($c/R$) = 2 km s$^{-1}$ in the case of $R$ = 90000.
% Note that Gaussian FWHM ($c/R$) is about 2 km s$^{-1}$ in the case of $R$ = 90 000.

We used the Gray (1989) assumption that the macroturbulence depends only on the surface gravity. We thus use

\begin{equation}
    v_{\mathrm{macro}} = 4.3 - 0.67 ~\log g ~\rm km~ s^{-1}.
\end{equation}

\hskip -15pt
We know the value of $v_{\mathrm{M}}$ from the observed spectrum. So the rotational velocity can be evaluated from Equation (1).
%Takeda et al. (2008) confirmed that the trend of results is in agreement with de Medeiros \& Mayor (1999) and Massarotti et al. (2008).

In this work, we measured $v_{\mathrm{M}}$ to be 3.9 km s$^{-1}$ and estimated the rotational velocity equal to 1.3 km~s$^{-1}$.
In the literature, the $v_{\mathrm{rot}}$ sin($i$) values are taken from various sources obtained by different methods.
de Medeiros \& Mayor (1999) measured rotational velocities using the cross-correlation technique on spectra obtained from the CORAVEL spectrometer with an uncertainty of about 1.0 km s$^{-1}$.
Hekker \& Mel{\'e}ndez (2007) determined rotational velocities from FWHM measurements using the Coud\'{e} Auxiliary Telescope (CAT) at Lick observatory. Lastly, Massarotti et al. (2008) used the cross-correlation of the observed spectra against templates drawn from a library of synthetic spectra calculated by Kurucz (1992) for different stellar atmospheres, even though $\sigma$ Per was not calculated in the study.

As stated above, estimated rotational velocities depend critically on the adopted templates and methods used for the calibration of the line broadening. Thus it would be inappropriate to choose any one value as the ``correct'' one. Using these determinations, we can define a range of $v_{\mathrm{rot}}$ sin($i$) measurements of 1.0$-$2.21~km~s$^{-1}$. Based on the rotational velocities and the stellar radius, we derived the range of upper limits for the rotational period of

   $P_{rot} = 2 \pi$$R_{\star}$/[$v_{rot}$ sin($i$)] = 641$-$1288 days.\\

\hskip -15pt

\section{RADIAL VELOCITY VARIATIONS AND ORIGIN}
Evolved stars exhibit pulsations as well as surface activity, which result in low-amplitude RV variability on different time scales. Generally, short-term RV variations have been known to be the result of stellar pulsations (Hatzes \& Cochran 1998), whereas long-term (hundreds of days) variations with a low-amplitude may be caused by three kinds of phenomena: planetary companions, stellar oscillations, or rotational modulations.
To establish the origin of the RV variabilities for $\sigma$~Per, we examined 1) the orbital fit, 2) the stellar chromospheric activity, 3) the \emph{HIPPARCOS} photometry, and 4) the spectral line bisectors.

\subsection{ORBITAL SOLUTIONS}
\begin{figure}[t]
\centering \epsfxsize=8cm
\epsfbox{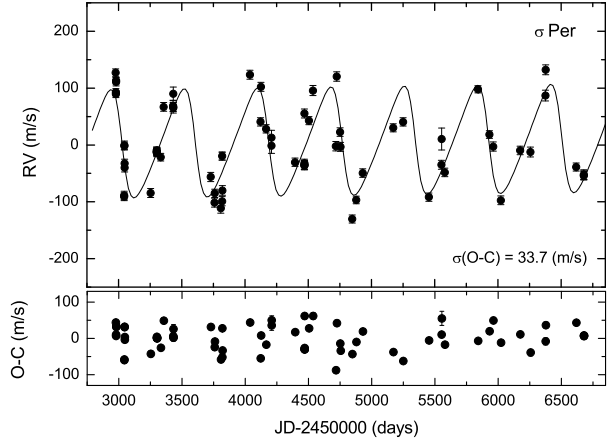}
\caption{RV measurements (top) and rms scatter of the residuals (bottom)
      for $\sigma$ Per taken at BOAO from December 2003 to January 2014.
      The solid line is the orbital solution with a period of 579.8 days
      and an eccentricity of 0.3.}
\label{orbit}
\end{figure}

\begin{figure}[!t]
\centering \epsfxsize=8cm
\epsfbox{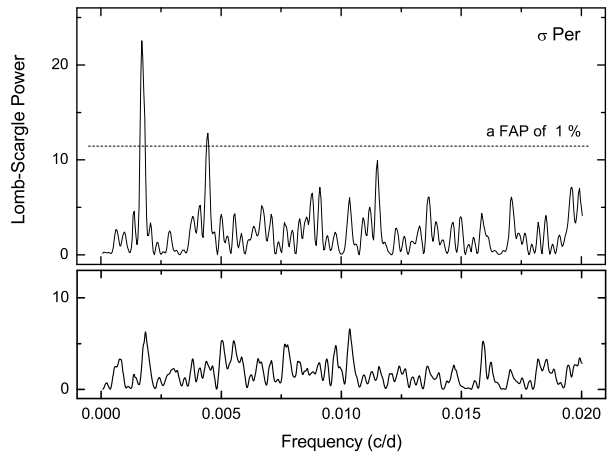}
\caption{The Lomb-Scargle periodogram of the RV measurements for $\sigma$~Per.
      The periogram shows a significant power peak at a frequency of 0.00172 c d$^{-1}$
      corresponding to a period of 579.8 days (top) and no significant power peak at the residual (bottom).}
\label{power}
\end{figure}

\begin{figure}[!t]
\centering \epsfxsize=8cm
\epsfbox{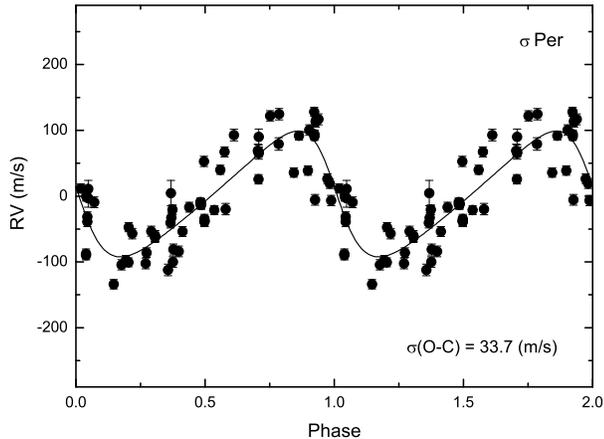}
\caption{Phase diagram for $\sigma$ Per.
    The solid line is phased with the orbital solution that fits the data  with a period of 579.8 days and an eccentricity of 0.3.
}
\label{phase}
\end{figure}

Period analysis was applied to assess the significance of periodic trends or variations of the time series. The Lomb-Scargle periodogram (Lomb 1976; Scargle 1982), which is a useful tool for investigating long-period variations for unequally spaced data, was used to calculate the RV time series. Our RV measurements for $\sigma$ Per are listed in Table \ref{tab1} and shown in Figure~\ref{orbit} (top panel).
The Lomb-Scargle periodogram for $\sigma$ Per shows a significant power peak at $f_{1}$=0.00172~c\,d$^{-1}$  and no significant peak in the residuals after subtracting the major period (Figure~\ref{power}). The best fit Keplerian orbit is found with to have a period $P$ = 579.8 $\pm$ 2.4 days, a semi-amplitude $K$ = 96.0 $\pm$ 6.5~m~s$^{-1}$, and an eccentricity $e$ = 0.3 $\pm$ 0.1. Assuming a stellar mass of 2.25 $\pm$ 0.5~$M_{\odot}$ (Luck \& Heiter 2007), which includes the range of mass (1.82 $M_{\odot}$) estimated by the photometric method, we derived a minimum mass of a planetary companion $m$~sin($i$) = 6.5 $\pm$ 1.0  $M_{\rm{Jup}}$ at a distance $a$ = 1.8 AU from the host. The RV phase diagram for the orbit is shown in Figure~\ref{phase}. All the orbital elements are listed in Table~ \ref{tab3}.

The statistical significance of this signal was calculated using the bootstrap randomization technique (K{\"u}rster et al. 1999). The measured RV values were randomly shuffled keeping the observed times fixed. The Lomb-Scargle periodogram was then computed for 200000 ``random'' datasets. The fraction of periodograms having power higher than the amplitude peak in the range of 0 $<$ $f$ $<$ 0.02 c d$^{-1}$ represents the false-alarm probability (FAP) that noise would create the detected signal. A small FAP is therefore an indication of a significant peak.
We found that among 200000 trials the highest peak of the randomly generated periodograms never exceeded that in the original data, thus the FAP is
$<$  $10^{-5}$. We could not find any significant power peaks in the residual periodogram (FAP of 4 $\times$ $10^{-2}$).

%-------------------------------------------------------------
\begin{table}[t]
\begin{center}
\caption{Orbital parameters for $\sigma$ Per b.}
\label{tab3}
\doublerulesep2.0pt
\renewcommand\arraystretch{1.2}
\begin{tabular}{lcc}
\hline
\hline
    Parameter                    & Unit    & Value                  \\

\hline
    Period                       &[days] & 579.8  $\pm$ 2.4       \\
    $\it T$$_{\rm{periastron}}$  &[JD]   & 2453022.6 $\pm$ 20.7   \\
    $\it{K}$                     &[m s$^{-1}$] & 96.0  $\pm$ 6.5  \\
    $\it{e}$                     & & 0.3   $\pm$ 0.1     \\
    $\omega$                     &[deg] & 83.5   $\pm$ 14.2       \\
    slope           &[m s$^{-1}$ day$^{-1}$] & $-$2.6 $\times$ 10$^{- 6}$  \\
    $\sigma$ (O-C)               &[m s$^{-1}$]  & 33.7               \\
\hline
    $\textit{$M_{\star}$}$       &[$M_{\odot}$]   & 2.25 $\pm$ 0.5   \\
    $m$ sin($i$)                 &[$M_{\rm{Jup}}$]& 6.5 $\pm$ 1.0    \\
    $\it{a}$                     &[AU]            & 1.8 $\pm$ 0.1    \\
\hline

\end{tabular}
\end{center}
\end{table}

\subsection{CHROMOSPHERIC ACTIVITIES}
The EW variations of Ca II H \& K, H$_{\alpha}$, and Ca II 8662 {\AA} lines are frequently
used a chromospheric activity indicators because they are sensitive to the stellar atmospheric activity.
Such activity could have a decisive effect on the RV variations.
While the H$_{\alpha}$ is sensitive to the atmospheric stellar activity (K{\"u}rster et al. 2003) and useful to measure the
variations, it is not easy to estimate the H$_{\alpha}$ EW due to line blending in stellar spectra and telluric
lines if any. Another activity indicator, the Ca II 8662~{\AA} line, is also not suitable because
significant fringing and saturation of our CCD spectra is seen at wavelengths longer than 7500~{\AA}.

\begin{figure}[!t]
\centering \epsfxsize=8cm
\epsfbox{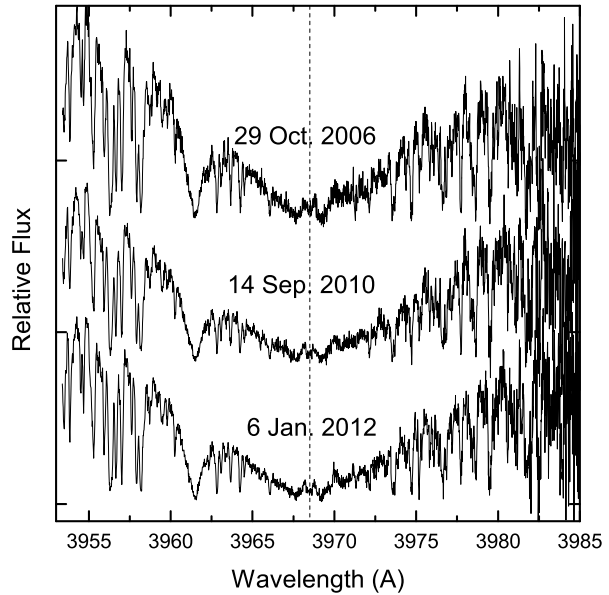}
\caption{The Ca II H spectral region for $\sigma$~Per at different phases. The Ca II H core features exhibit weak emission at the line center. Line profiles do not show any significantly different features related to the RV variations.
}
\label{caII}
\end{figure}

The emission of the Ca~II H \& K comes from the chromosphere and the core of the spectral line often exhibits central
reversal in the presence of chromospheric activity (Saar \& Donahue 1997). It is common in cool
stars and is intimately connected to the convective envelope and magnetic activity. Unfortunately,
the spectra are not clear enough to resolve the emission feature in the Ca~II~K line core for
$\sigma$ Per. Instead, we use the relatively stronger Ca~II~H region shown in Figure~\ref{caII}.
There are weak central emission lines with weak reversal in the core of Ca~II~H.
We used the averaged Ca~II~H line profiles at the upper, lower, and near zero part of the RV curve to
identify any systematic difference related to the RV variations. For $\sigma$~Per,
these correspond to the positive deviation on JD-2454038.348295 (RV = 123.6 m s$^{-1}$), the negative
deviation on JD-2455454.324212 (RV = $-$ 91.8~m~s$^{-1}$), and the zero deviation on
JD-2455932.979094 (RV = 18.3 m s$^{-1}$) of the RV curve.
Core emissions and reversals in the Ca II H central region are very weak and there are no systematic
differences in the line profiles at these three extremal parts of the RV curves. This means that
$\sigma$~Per, at the moment of observations, exhibited at most a modest chromospheric activity.

\subsection{PHOTOMETRIC VARIATIONS}
\begin{figure}[!t]
   \centering
   \includegraphics[width=8cm]{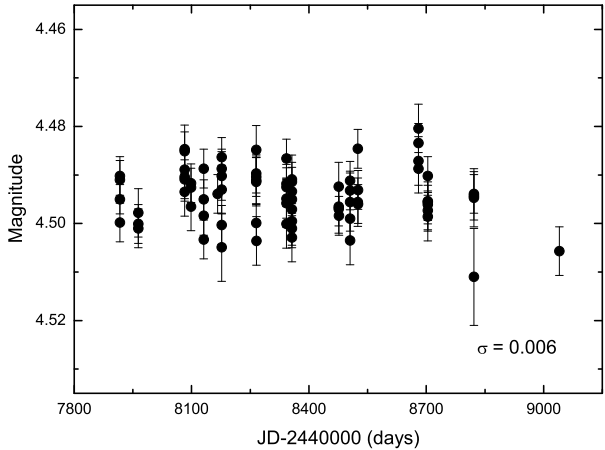}
      \caption{The \emph{HIPPARCOS} photometric variations from January 1990 to February 1993 for $\sigma$ Per.
        }
        \label{hip}
   \end{figure}

We analyzed the \emph{HIPPARCOS} photometry data to search for possible brightness variations which
might be caused by the rotational modulation of cool stellar spots. The available photometry database
comprises 77 \emph{HIPPARCOS} measurements for $\sigma$ Per from January 1990 to February 1993.
They maintained photometric stability down to rms scatters of 0.006 magnitude, which correspond to
0.14\% variations over the time span of the observations. Figure~\ref{hip} shows  no significant photometric variations.

%
%-----------------------------------------------------------

\subsection{LINE BISECTOR}
The RV variations due to planetary companions should not produce any changes in the spectral
line shape whereas the surface inhomogeneities do. Stellar rotational modulations of
surface inhomogeneities can create variable asymmetries in the spectral line profiles. Thus, the analysis
of the shapes of the spectral lines, in particular the analysis of the line bisectors, may identify
the origin of observed RV variations  and prove or disprove the existence of planetary
companions (e.g., Queloz et al. 2001).

\begin{figure}[!t]
   \centering
   \includegraphics[width=8cm]{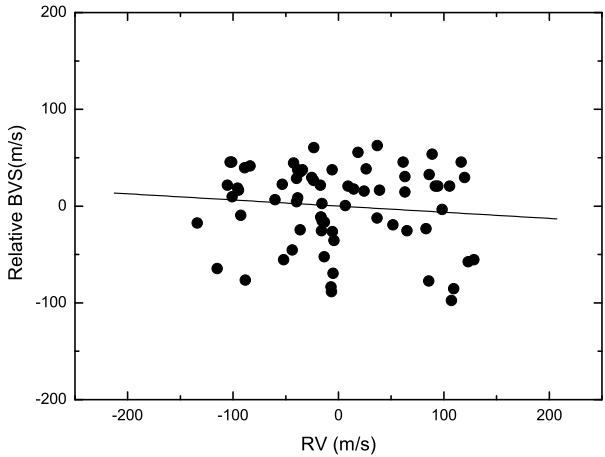}
      \caption{Relation between the RV and BVS variations for the Ni I 6643.6 {\AA} spectral line. The solid line marks the slope of the spectral line.
        }
        \label{bvs}
   \end{figure}

The RV differences of the central values at high and low flux levels of the line profiles are
defined as a bisector velocity span (BVS). To calculate the BVS, following Hatzes et al. (2005)
and Lee et al. (2013), we selected the unblended strong line of Ni I 6643.6 {\AA} that is located beyond
the I$_{2}$ absorption region. We estimated the BVS of the profile between two different flux levels
at 0.8 and 0.4 of the central depth as the span points, which avoided the spectral core and wings
where the differences in the bisector measurements are large and noisy. The BVS variations for $\sigma$ Per as
a function of RV are shown in Figure~\ref{bvs}. There is no obvious correlation between the RV and
BVS measurements.

%-----------------------------------------------------------

\section{DISCUSSION}
Our analysis of the RV measurements for $\sigma$~Per revealed a 579 day periodic variation that
persisted between 2003 and 2014 for almost six and half cycles. In evolved stars, low-amplitude and
long-term RV variations, such as in hundreds of days, may be caused by stellar pulsations,
rotational modulations of inhomogeneous surface features, or
planetary companions. Of these, RV variations due to pulsations are unlikely because the period of
the fundamental radial mode pulsation is of several days and two orders of magnitude longer compared to
observed RV variations.
The Ca II H  line was used to monitor chromospheric activities, but does not show any significant
variations. The \emph{HIPPARCOS} photometry does not show any detectable variations nor the
BVSs show any correlation with the RV measurements. All these null results from the analysis
of the chromospheric activities or surface inhomogeneities argue against the rotational modulation
of surface activities as the origin of observed RV variations and in favor of the planetary companion
hypothesis.

Our precise RV measurements for $\sigma$~Per have established the presence of variations with a
579.8~$\pm$~2.4 day period. Due to the uncertainty in the mass of the host star, the planetary companion mass
can range from  5.5$-$7.5 $M_{\rm{Jup}}$ with an orbital semi-major axis of 1.8~$\pm$~0.1 AU, and
an eccentricity of 0.3 $\pm$ 0.1. The planetary companion to $\sigma$~Per has similar properties to
other companions around K giant stars. They are mostly massive planets with the average mass of
$\sim$~5~$M_{\rm{Jup}}$ and an orbital radius of $\sim$ 1.5 AU. The low metallicity of $-$ 0.2 for
$\sigma$~Per is also consistent with the average value estimated for other planet-hosting stars
($\rm{[Fe/H]}$ $\sim$ $-$ 0.12).

The residual RV variations after removing the 579.8-day signal are 33.7 m s$^{-1}$.
This value of scatter is  significantly greater than the RV precision for the RV standard star
$\tau$ Ceti (6.8 m s$^{-1}$) and the typical internal error of individual RV
accuracies for K giants from our survey of $\sim$ 7.8 m s$^{-1}$.
The significant residual appears to be caused by systematics in RV measurement.
Otherwise, the variations, whether periodic or not, are typical in evolved K giants (Setiawan et al. 2003; Hatzes et al. 2005;
D{\"o}linger et al. 2007; de Medeiros et al. 2009; Han et al. 2010; Lee et al. 2012c) and can be possibly attributed to
pulsational and convection activity. Furthermore, residual RV variations tend to increase toward later spectral types.

%--------------------------------------------------------------------
\acknowledgments{
BCL acknowledges partial support by the KASI (Korea Astronomy and Space Science Institute) grant
2014-1-400-06. MGP was supported by Basic Science Research Program through the National Research
Foundation of Korea(NRF) funded by the Ministry of Education, Science and Technology (2012R1A1A4A010
13596). DEM acknowledges his work as part of the research activity of the National Astronomical
Research Institute of Thailand (NARIT), which is supported by the Ministry of Science and Technology
of Thailand. This research made use of the SIMBAD database, operated at the CDS, Strasbourg, France.
We thank the developers of the Bohyunsan Observatory Echelle Spectrograph (BOES) and all staff of
the Bohyunsan Optical Astronomy Observatory (BOAO). }

%--------------------------------------------------------------------

%-------------------------------------------------------------------

\begin{thebibliography}{}

\bibitem[Bertelli et al.(1994)]{1994A&AS..106..275B}
    Bertelli, G., Bressan, A., Chiosi, C., Fagotto, F., \& Nasi, E.\ 1994,
    Theoretical isochrones from models with new radiative opacities,
    \aaps, 106, 275

\bibitem[Butler et al.(1996)]{1996PASP..108..500B}
    Butler, R.~P., Marcy, G.~W., Williams, E., McCarthy, C., Dosanjh, P., Vogt, S.~S. \ 1996,
    Attaining Doppler Precision of 3 m s$^{-1}$,
    \pasp, 108, 500

\bibitem[de Medeiros \& Mayor(1999)]{1999A&AS..139..433D}
    de Medeiros, J.~R., \& Mayor, M.\ 1999,
    A catalog of rotational and radial velocities for evolved stars,
    \aaps, 139, 433

\bibitem[de Medeiros et al.(2009)]{2009A&A...504..617D}
    de Medeiros, J.~R., Setiawan, J., Hatzes, A.~P., Pasquini, L., Girardi, L., Udry, S., D{\"o}llinger, M.~P., \& da Silva, L. \ 2009,
    A planet around the evolved intermediate-mass star HD 110014,
    \aap, 504, 617

\bibitem[D{\"o}llinger et al.(2007)]{2007A&A...472..649D}
    D{\"o}llinger, M.~P., Hatzes, A.~P., Pasquini, L., Guenther, E.~W., Hartmann, M., Girardi, L., \& Esposito, M. \ 2007,
    Discovery of a planet around the K giant star 4 Ursae Majoris,
    \aap, 472, 649

\bibitem[Endl et al.(2000)]{2000A&A...362..585E}
    Endl, M., K{\"u}rster, M., \& Els, S.\ 2000,
    The planet search program at the ESO Coud{\'e} Echelle spectrometer. I. Data modeling technique and radial velocity precision tests,
    \aap, 362, 585

\bibitem[Frink et al.(2002)]{2002ApJ...576..478F}
    Frink, S., Mitchell, D.~S., Quirrenbach, A., Fischer, D.~A., Marcy, G.~W., \& Butler, R.~P. \ 2002,
    Discovery of a Substellar Companion to the K2 III Giant {$\iota$} Draconis,
    \apj, 576, 478

\bibitem[Galazutdinov (1992)]{}
    Galazutdinov, G. A. 1992,
    Special Astrophysical Observatory Preprint 92 (Nizhnij Arkhyz: SAO)

\bibitem[Gray(1989)]{1989ApJ...347.1021G}
    Gray, D.~F.\ 1989,
    The rotational break for G giants,
    \apj, 347, 1021

\bibitem[Han et al.(2007)]{2007PKAS...22...75H}
    Han, I., Kim, K.-M., Byeong-Cheol, L., \& Valyavin, G.\ 2007,
    Development of RVI2CELL -A Precise Radial Velocity Estimation Program with BOES Data,
    Publication of Korean Astronomical Society, 22, 75

\bibitem[Han et al.(2008)]{2008JKAS...41...59H}
    Han, I., Lee, B.-C., Kim, K.-M., \& Mkrtichian, D.~E.\ 2008,
    Confirmation of the Exoplanet around {$\beta$} GEM from the RV Observations Using BOES,
    Journal of Korean Astronomical Society, 41, 59

\bibitem[Han et al.(2010)]{2010A&A...509A..24H}
    Han, I., Lee, B.-C., Kim, K.~M., Mkrtichian, D.~E., Hatzes, A.~P., \& Valyavin, G. \ 2010,
    Detection of a planetary companion around the giant star {$\gamma$}$^{1}$ Leonis,
    \aap, 509, A24

\bibitem[Hatzes \& Cochran(1998)]{1998ASPC..154..311H}
    Hatzes, A.~P., \& Cochran, W.~D.\ 1998,
    Stellar Oscillations in K Giant Stars,
    Cool Stars, Stellar Systems, and the Sun, 154, 311

\bibitem[Hatzes et al.(2005)]{2005A&A...437..743H}
    Hatzes, A.~P., Guenther, E.~W., Endl, M., Cochran, W.~D., D{\"o}llinger, M.~P., \& Bedalov, A. \ 2005,
    A giant planet around the massive giant star HD 13189,
    \aap, 437, 743

\bibitem[Hekker \& Mel{\'e}ndez(2007)]{2007A&A...475.1003H}
    Hekker, S., \& Mel{\'e}ndez, J.\ 2007,
    Precise radial velocities of giant stars. III. Spectroscopic stellar parameters,
    \aap, 475, 1003

\bibitem[Kim et al.(2007)]{2007PASP..119.1052K}
    Kim, K.-M., Han, I., Valyavin, G.~G., Plachinda, S., Jang, J.~G., Jang, B.-H., Seong, H.~C., Lee, B.-C., Kang, D.-I., Park, B.-G., Yoon, T.~S., \& Vogt, S.~S. \ 2007,
    The BOES Spectropolarimeter for Zeeman Measurements of Stellar Magnetic Fields,
    \pasp, 119, 1052

\bibitem[K{\"u}rster et al.(1999)]{1999A&A...344L...5K}
    K{\"u}rster, M., Hatzes, A.~P., Cochran, W.~D., et al.\ 1999,
    Precise radial velocities of Proxima Centauri. Strong constraints on a substellar companion,
    \aap, 344, L5

\bibitem[K{\"u}rster et al.(2003)]{2003A&A...403.1077K}
    K{\"u}rster, M., Endl, M., Rouesnel, F., et al.\ 2003,
    The low-level radial velocity variability   in Barnard's star (= GJ 699).  Secular acceleration, indications for convective redshift,  and planet mass limits,
    \aap, 403, 1077

\bibitem[Kurucz(1992)]{1992IAUS..149..225K}
    Kurucz, R.~L.\ 1992,
    Model Atmospheres for Population Synthesis,
    The Stellar Populations of Galaxies, 149, 225

\bibitem[Lee et al.(2008)]{2008AJ....135.2240L}
    Lee, B.-C., Mkrtichian, D.~E., Han, I., Park, M.-G., \& Kim, K.-M.\ 2008,
    Precise Radial Velocities of Polaris: Detection of Amplitude Growth,
    \aj, 135, 2240

\bibitem[Lee et al.(2011)]{2011A&A...529A.134L}
    Lee, B.-C., Mkrtichian, D.~E., Han, I., Kim, K.-M., \& Park, M.-G.\ 2011,
    A likely exoplanet orbiting the oscillating K-giant {$\alpha$} Arietis,
    \aap, 529, A134

\bibitem[Lee et al.(2012)]{2012A&A...543A..37L}
    Lee, B.-C., Han, I., Park, M.-G., Kim, K.-M., \& Mkrtichian, D.~E.\ 2012a,
    Detection of the 128-day radial velocity variations in the supergiant {$\alpha$} Persei. Rotational modulations, pulsations, or a planet?,
    \aap, 543, A37

\bibitem[Lee et al.(2012)]{2012A&A...546A...5L}
    Lee, B.-C., Han, I., Park, M.-G., Mkrtichian, D.~E., \& Kim, K.-M.\ 2012b,
    A planetary companion around the K giant {$\epsilon$} Corona Borealis,
    \aap, 546, A5

\bibitem[Lee et al.(2012)]{2012A&A...548A.118L}
    Lee, B.-C., Mkrtichian, D.~E., Han, I., Park, M.-G., \& Kim, K.-M.\ 2012c,
    Detection of an exoplanet around the evolved K giant HD 66141,
    \aap, 548, A118

\bibitem[Lee et al.(2013)]{2013A&A...549A...2L}
    Lee, B.-C., Han, I., \& Park, M.-G.\ 2013,
    Planetary companions orbiting M giants HD 208527 and HD 220074,
    \aap, 549, A2

\bibitem[Lomb(1976)]{1976Ap&SS..39..447L}
    Lomb, N.~R.\ 1976,
    Least-squares frequency analysis of unequally spaced data,
    \apss, 39, 447

\bibitem[Luck \& Heiter(2007)]{Luck07}
    Luck, R.~E., \& Heiter, U.\ 2007,
    Giants in the Local Region,
    \aj, 133, 2464

\bibitem[Massarotti et al.(2008)]{2008AJ....135..209M}
    Massarotti, A., Latham, D.~W., Stefanik, R.~P., \& Fogel, J.\ 2008,
    Rotational and Radial Velocities for a Sample of 761 HIPPARCOS Giants and the Role of Binarity,
    \aj, 135, 209

\bibitem[Mayor \& Queloz(1995)]{1995Natur.378..355M}
    Mayor, M., \& Queloz, D.\ 1995,
    A Jupiter-mass companion to a solar-type star,
    \nat, 378, 355

\bibitem[McDonald et al.(2012)]{mc12}
    McDonald, I., Zijlstra, A.~A., \& Boyer, M.~L.\ 2012,
    Fundamental parameters and infrared excesses of Hipparcos stars,
    \mnras, 427, 343

\bibitem[Pasinetti-Fracassini et al.(2001)]{pas01}
    Pasinetti-Fracassini, L.~E., Pastori, L., Covino, S., \& Pozzi, A.\ 2001,
    Catalogue of Apparent Diameters and Absolute Radii of Stars (CADARS) - Third edition - Comments and statistics,
    \aap, 367, 521

\bibitem[Queloz et al.(2001)]{2001A&A...379..279Q}
    Queloz, D., Henry, G.~W., Sivan, J.~P., et al.\ 2001,
    No planet for HD 166435,
    \aap, 379, 279

\bibitem[Saar \& Donahue(1997)]{1997ApJ...485..319S}
    Saar, S.~H., \& Donahue, R.~A.\ 1997,
    Activity-related Radial Velocity Variation in Cool Stars,
    \apj, 485, 319

\bibitem[Scargle(1982)]{1982ApJ...263..835S}
    Scargle, J.~D.\ 1982,
    Studies in astronomical time series analysis. II - Statistical aspects of spectral analysis of unevenly spaced data,
    \apj, 263, 835

\bibitem[Setiawan et al.(2003)]{2003A&A...397.1151S}
    Setiawan, J., Pasquini, L., da Silva, L., von der L{\"u}he, O., \& Hatzes, A.\ 2003,
    Precise radial velocity measurements of G and K giants.  First results,
    \aap, 397, 1151

\bibitem[Takeda et al.(2005)]{2005PASJ...57...27T}
    Takeda, Y., Ohkubo, M., Sato, B., Kambe, E., \& Sadakane, K.\ 2005,
    Spectroscopic Study on the Atmospheric Parameters of Nearby F--K Dwarfs and Subgiants,
    \pasj, 57, 27

\bibitem[Takeda et al.(2008)]{2008PASJ...60..781T}
    Takeda, Y., Sato, B., \& Murata, D.\ 2008,
    Stellar Parameters and Elemental Abundances of Late-G Giants,
    \pasj, 60, 781

\bibitem[Valenti et al.(1995)]{1995PASP..107..966V}
    Valenti, J.~A., Butler, R.~P., \& Marcy, G.~W.\ 1995,
    Determining Spectrometer Instrumental Profiles Using FTS Reference Spectra,
    \pasp, 107, 966

\bibitem[van Leeuwen(2007)]{2007A&A...474..653V}
    van Leeuwen, F.\ 2007,
    Validation of the new Hipparcos reduction,
    \aap, 474, 653


\end{thebibliography}
\end{document}